\documentclass[aps, preprint, superscriptaddress, onecolumn, longbibliography]{revtex4-1}

\usepackage{xcolor}
\definecolor{mygreen}{RGB}{0,128,0}
\usepackage{graphicx}
\usepackage{latexsym}
\usepackage{amssymb}
\usepackage{amsmath}
\usepackage{amsfonts}
\usepackage{units}
\usepackage{upgreek}
\usepackage{bm}
\usepackage{multirow}
\usepackage{enumitem}
\usepackage{hyperref}
\usepackage{lipsum}
\usepackage{tabularx}
\usepackage{hhline}
\usepackage[version=4]{mhchem} % Formula subscripts using \ce{}
\usepackage{verbatim}

\begin{document}

\title{Phase discovery with active learning:\\ Application to structural phase transitions in equiatomic NiTi}

\author{Jonathan Vandermause}
%\email{jonathan\_vandermause@g.harvard.edu}
\affiliation{Department of Physics, Harvard University, Cambridge, Massachusetts 02138, USA}
\affiliation{John A. Paulson School of Engineering and Applied Sciences, Harvard University, Cambridge, MA 02138, USA}

\author{Anders Johansson}
\affiliation{John A. Paulson School of Engineering and Applied Sciences, Harvard University, Cambridge, MA 02138, USA}

\author{Yucong Miao}
\affiliation{John A. Paulson School of Engineering and Applied Sciences, Harvard University, Cambridge, MA 02138, USA}

\author{Joost J. Vlassak}
\affiliation{John A. Paulson School of Engineering and Applied Sciences, Harvard University, Cambridge, MA 02138, USA}

\author{Boris Kozinsky}
\email{bkoz@seas.harvard.edu}
\affiliation{John A. Paulson School of Engineering and Applied Sciences, Harvard University, Cambridge, MA 02138, USA}
\affiliation{Robert Bosch LLC, Research and Technology Center, Cambridge, MA 02139, USA}

\date{\today}

\begin{abstract}
Nickel titanium (NiTi) is a protypical shape-memory alloy used in a range of biomedical and engineering devices, but direct molecular dynamics simulations of the martensitic $B19' \rightarrow B2$ phase transition driving its shape-memory behavior are rare and have relied on classical force fields with limited accuracy. Here, we train four machine-learned force fields for equiatomic NiTi based on the LDA, PBE, PBEsol, and SCAN DFT functionals. The models are trained on the fly during NPT molecular dynamics, with DFT calculations and model updates performed automatically whenever the uncertainty of a local energy prediction exceeds a chosen threshold. The models achieve accuracies of 1-2 meV/atom during training and are shown to closely track DFT predictions of $B2$ and $B19'$ elastic constants and phonon frequencies. Surprisingly, in large-scale molecular dynamics simulations, only the SCAN model predicts a reversible $B19' \leftrightarrow B2$ phase transition, with the LDA, PBE, and PBEsol models predicting a reversible transition to a previously uncharacterized low-volume phase, which we hypothesize to be a new stable high-pressure phase. We examine the structure of the new phase and estimate its stability on the temperature-pressure phase diagram. This work establishes an automated active learning protocol for studying displacive transformations, reveals important differences between DFT functionals that can only be detected in large-scale simulations, provides an accurate force field for NiTi, and identifies a new phase.

\end{abstract}

\maketitle

\section{Introduction}
The bimetallic alloy nickel titanium (NiTi) exhibits a range of fascinating behaviors, including superelasticity and the shape-memory effect. Because of its unique properties, NiTi is used in a range of biomedical, aerospace, and robotics applications \cite{jani2014review}. Its shape-memory behavior is driven by a structural phase transition between a high-temperature $B2$ austenite phase and a low-temperature $B19'$ martensite phase at $\sim 340 \text{ K}$ \cite{frenzel2010influence}. Direct molecular dynamics simulations of the martensitic phase transition in NiTi could shed light on the microscopic mechanisms of the shape-memory effect and even guide the rational design of shape-memory alloys with tunable properties, but an accurate atomistic description of the stable phases of equiatomic NiTi has posed a number of theoretical challenges over the last two decades. The principal computational tool available for studying the microsopic properties of NiTi, density functional theory (DFT), has revealed a complex picture of its potential energy surface, with several distinct phases lying close in energy to the convex hull of the phase diagram \cite{huang2001lattice, huang2003crystal, holec2011ab, wang2012resolving, jain2013commentary, haskins2016ab}. For example, DFT predicts the minimum-energy phase at zero Kelvin to be the orthorombic $B33$ phase (space group $C_{mcm}$) rather than the $B19'$ phase (space group $P2_1/m$) reported in experiments, with the energy gap between the two as small as $8 \text{ meV/atom}$ \cite{huang2003crystal}. A recent \textit{ab initio} molecular dynamics study demonstrated that $B19'$ has lower free energy than $B33$ above $50 \text{ K}$, providing evidence that $B19'$ is stabilized by entropic effects \cite{haskins2016ab}. However, there has to date been no direct \textit{ab initio} simulation of the martensitic phase transition, due to the extreme computational cost that a sufficiently large simulation would require.

A more computationally feasible approach to simulating the martensitic phase transition is to use an empirical interatomic potential or force field to drive the simulation, as the cost of energy and force prediction with a classical force field scales only linearly with the size of the system. However, parameterizing an accurate classical force field that captures the rich and subtle potential energy surface of NiTi is a formidable challenge. A number of classical force fields for NiTi have been developed over the years \cite{farkas1996atomistic, lai2000lattice, ishida2007md, mutter2010simulation, saitoh2010atomic, zhong2011atomistic, ko2015development, kim2017development, kavousi2019modified}, with the 2NN MEAM model of \cite{ko2015development} specifically targeting the martensitic phase transition. These classical models rely on strict functional forms, with a fairly small number of parameters fit to a manually constructed DFT database or a small set of experimentally measured properties.

In the last decade, machine learning (ML) based approaches have dramatically increased the flexibility of classical force fields, opening up the possibility of linear-scaling models that achieve very high accuracy relative to their \textit{ab initio} reference data \cite{behler2007generalized, handley2009optimal, bartok2010gaussian, thompson2015spectral, wood2018extending}. A highly accurate ML force field for NiTi would help address the fundamental challenge of simulating the martensitic phase transition at \textit{ab initio} accuracy, offering both quantum accuracy and efficiency of a classical force field. While many early ML force fields were restricted to relatively simple single-element systems \cite{szlachta2014accuracy, bartok2018machine, deringer2020general}, recently a neural network based force field was developed for NiTi based on the PBE DFT functional \cite{tang2022high}. However, a reversible martensitic phase transition was reported with this force field only in a very small 6x4x4 supercell, with manually selected phases included in the training set and stress labels excluded.

Here, we train four highly accurate ML force fields for equiatomic NiTi based on the LDA, PBE, PBEsol, and SCAN DFT functionals. We use our recently developed active learning methodology \cite{vandermause2020fly, xie2021bayesian, vandermause2021active} to autonomously train the models ``on the fly'' during molecular dynamics. Structural phase transitions are directly observed during training, with all models predicting the $B19' \rightarrow B2$ transition on heating at zero pressure. On cooling, however, the LDA, PBE, and PBEsol models predict transitions from $B2$ to a previously uncharacterized phase, which we call $M2$. Only the SCAN model predicts the martensitic $B2 \rightarrow B19'$ phase transition. The results are confirmed in much larger cooling/heating simulations of $12,168$ atoms, which show reversible $B2 \rightarrow M2$ predicted by the LDA, PBE, and PBEsol models and a reversible $B2 \rightarrow B19'$ predicted by the SCAN model. We hypothesize that the new $M2$ phase is stable at high pressure.

\section{Results}
\subsection{On-the-fly training of NiTi force fields}
Fig.\ 1 illustrates our autonomous active learning approach to training NiTi force fields during molecular dynamics simulations. We train many-body sparse Gaussian process (SGP) models on four DFT functionals: LDA, PBE, PBEsol, and SCAN (see Sec.\ A, Methods for details). To sample different phases of NiTi in an unbiased fashion, we perform two independent $50$-ps on-the-fly training simulations of $144$-atom NiTi structures in the NPT ensemble at zero pressure, i.e.\ with cell lengths and angles allowed to freely change. The first is a cooling simulation that begins in the $B2$ austenite phase at $1000 \text{ K}$ and terminates at $50 \text{ K}$ (blue lines, Fig.\ 1(b-f)); the second is a heating simulation that begins in the $B19'$ martensite phase at low temperature and terminates at $1500 \text{ K}$ (red lines, Fig.\ 1(b-f)). In both the cooling and heating simulations, temperature shocks are introduced every $5$ ps during the first half of the simulation (vertical lines, Fig.\ 1(b-f)) to induce structural phase transitions during training. Example snapshots from the beginning and end of both simulations are shown in Fig.\ 1(a, d) for the PBE and SCAN models, illustrating the changes in cell shape and size observed during the simulations.

\begin{figure*}
	\centering
    \includegraphics{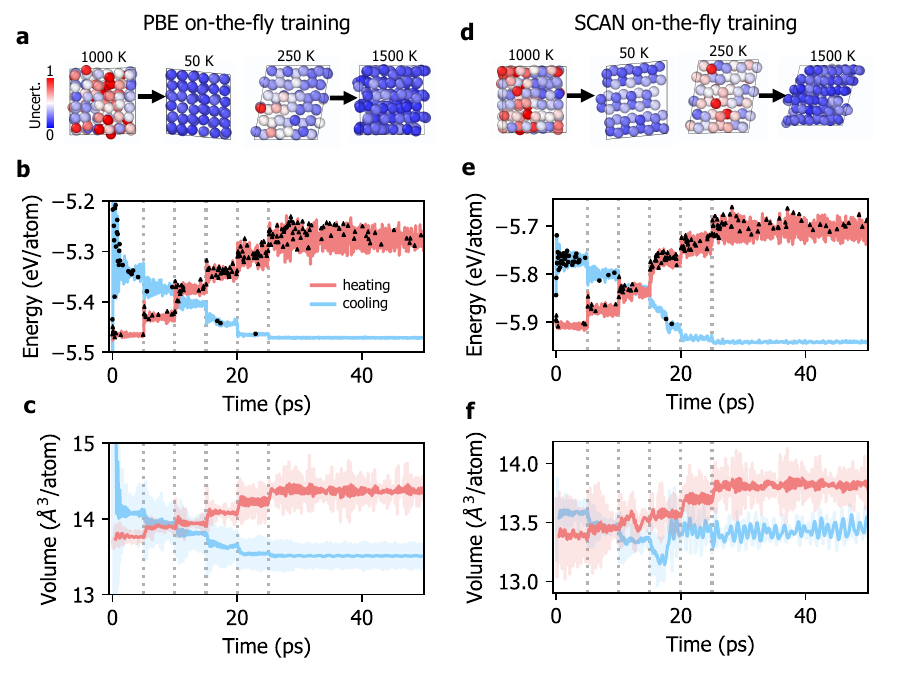}
	\caption{On-the-fly training of NiTi force fields. (a, d) Snapshots from PBE and SCAN training simulations. Atoms are colored by uncertainty in the local energy prediction. (b, e) Potential energy predictions during the cooling and heating simulations. DFT potential energies are shown in black, with vertical lines indicating changes in temperature. (c, f) Atomic volumes during training. For clarity, a moving average is plotted in dark red and blue.}
	\label{fig1}
\end{figure*}

Our training protocol is based on our previously introduced active learning method \cite{vandermause2020fly, vandermause2021active}, with individual MD steps driven by a Bayesian SGP model that is automatically updated when high-uncertainty local environments are encountered in the simulation (see Sec.\ B, Methods for details). In each training simulation, the model starts from scratch, with no data initially stored in the training set. DFT calculations (shown in black in Fig.\ 1(b,e)) are performed whenever the uncertainty of a local energy prediction exceeds a fixed prediction threshold. When a DFT calculation is performed, the training set of the model is updated with the potential energy, atomic forces and virial stress tensor obtained from the calculation, and subsequent MD steps are taken with the refined model. In general, the models display excellent agreement with DFT during training, with typical errors around $ 1 \text{ meV/atom}$ for energies, $50 \text{ meV/\AA}$ for forces, and $0.1 \text{ GPa}$ for stresses in all training simulations (see Figs.\ S5-S8).

Structural phase transitions are observed in all zero pressure training simulations, with heating simulations displaying the expected $B19' \rightarrow B2$ transition. This is most clearly seen in plots of the cell lengths and angles as a function of time (Figs.\ S1-S4), with sharp changes in the cell observed between $t = 15$ and $t = 20 \text{ ps}$ as the system is heated above $1000 \text{ K}$. Surprisingly, in the cooling simulations, the expected $B2 \rightarrow B19'$ was observed with the SCAN model only, as indicated by the sharp increase in atomic volume (blue line, Fig.\ 1(f)) and the formation of a monoclinic distortion in the unit cell (Fig.\ S2) observed at $t \approx 18 \text{ ps}$. Structural phase transitions are also observed with the PBE, LDA, and PBEsol models during cooling, but to lower-volume structures that are inconsistent with the expected $B19'$ martensite phase.

Visualizations of the structures obtained in each of the four cooling simulations are presented in Fig.\ S9. The structures are drawn from the final picosecond of the simulations, and subsequently relaxed with the final trained models to eliminate residual forces and stresses. The relaxed SCAN structure has space group $P2_1/m$ and exhibits a monoclinic distortion and elongated $\vec{c}$ lattice vector, consistent with $B19'$ martensite (Fig.\ S9(b)). The PBE, LDA, and PBEsol relaxed structures, on the other hand, have a rhombic prism unit cell and exhibit the orthorombic space group $C_{mcm}$. Curiously, this is the same space group as the ground state $B33$ phase of NiTi first characterized in \cite{huang2003crystal}, but the 144-atom orthorombic structures we observe are not consistent with $B33$ as they cannot be reduced to a simple 4-atom primitive cell and have significantly smaller atomic volume. To our best knowledge, this low-volume phase has not been characterized before, but is the stable phase predicted by our LDA, PBE, and PBEsol models upon cooling of $B2$ NiTi below the transition temperature.
%Our heating simulations suggest that the LDA, PBE, and PBEsol models predict $B19'$ to be metastable until transforming to $B2$ at high $T$.

\begin{figure*}
	\centering
    \includegraphics{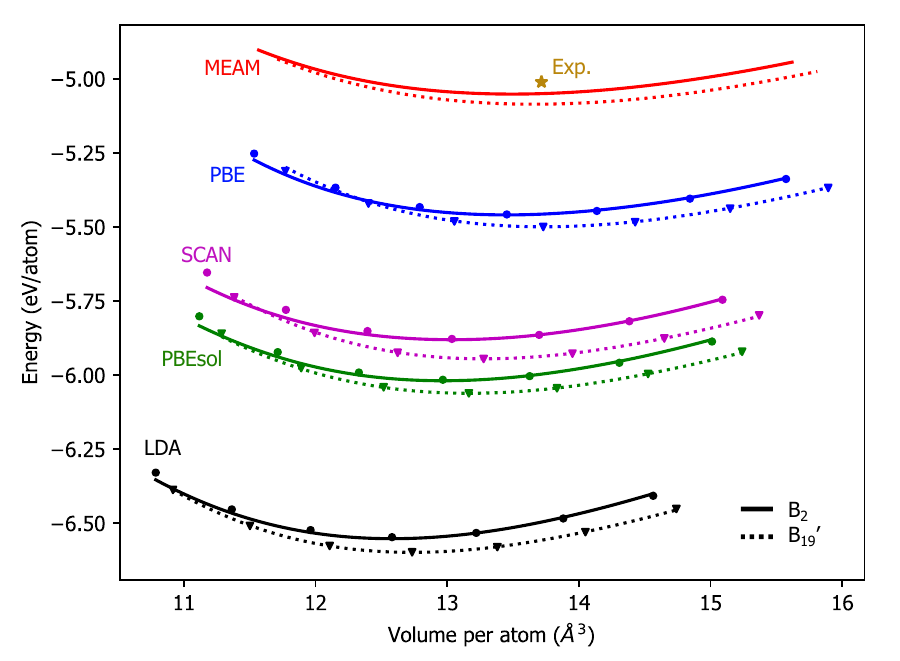}
	\caption{Cohesive energy versus volume of $B2$ and $B19'$ NiTi predicted by the four models trained in this work (LDA, black; PBEsol, green; PBE, blue; and SCAN, magenta) and by the MEAM model of Ref.\ \cite{ko2015development} (red). DFT cohesive energies are shown as dots for $B2$ and triangles for $B19'$. Experimental estimates of the cohesive energy and atomic volume of $B2$ NiTi derived from Refs.\ \cite{kubaschewski1956reaction, kittel1996introduction, prokoshkin2004lattice} are shown as a gold star.}
	\label{fig2}
\end{figure*}

\subsection{Model validation}
We validate our models by computing a range of challenging properties and comparing against DFT and a recent 2NN-MEAM NiTi force field \cite{ko2015development}. We first consider predictions of cohesive energies as a function of atomic volume for the $B2$ and $B19'$ phases (Fig.\ 2). The plotted volumes range between $86\%$ and $114\%$ of the equilibrium volumes and are obtained by applying isotropic strains to the fully relaxed $B2$ and $B19'$ structures predicted by each model. This allows us to examine model performance over a wide range of pressures between approximately $-20$ and $40 \text{ GPa}$ that were not encountered in the zero-pressure training simulations. DFT cohesive energies are plotted for a subset of these strained structures and show excellent agreement with our models over the entire range of volumes.  For reference, we include experimental estimates of the cohesive energy and atomic volume of $B2$ NiTi derived from Refs.\ \cite{kubaschewski1956reaction, kittel1996introduction, prokoshkin2004lattice} (gold star, Fig.\ 2). The four DFT functionals overestimate the cohesive energy of $B2$ and underestimate the atomic volume, with LDA showing the most severe disagreement with experiment. The MEAM force field is in the best agreement with experiment of the five force field models, in part because the experimentally measured cohesive energies of pure Ni and Ti were directly incorporated into the fit.

We next consider the elastic constants of $B2$ and $B19'$ NiTi. Predictions of the bulk modulus and three independent elastic constants of the $B2$ phase are plotted in Fig.\ S10, with generally good agreement between DFT and the machine-learned force fields, despite the absence of any relaxed or low-temperature $B2$ structures in the training set of the models. The MEAM model is in better agreement with experiment than DFT for the $C_{11}$ and $C_{12}$ constants, but overestimates $C_{44}$ by more than $130\%$.

\begin{figure*}
	\centering
    \includegraphics{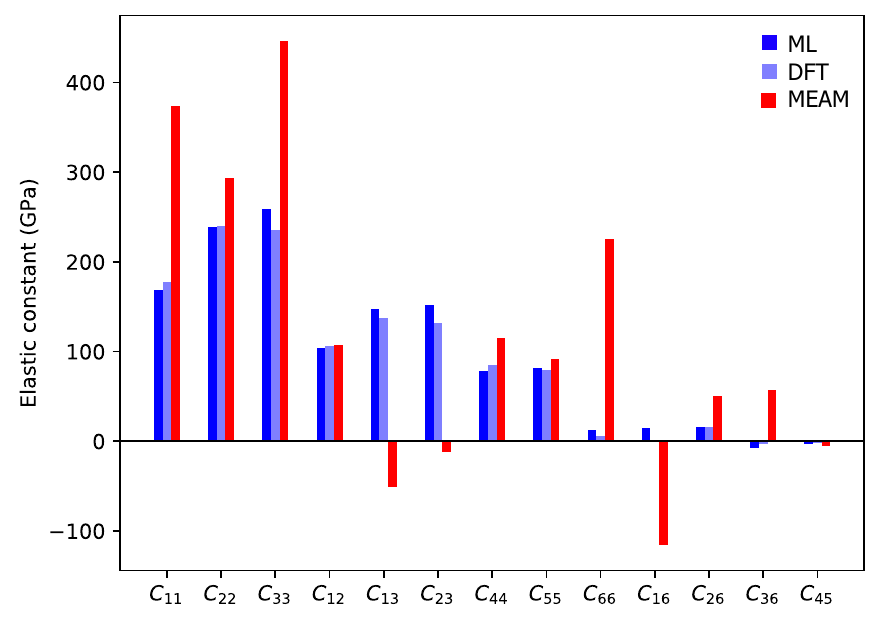}
	\caption{Elastic constants of $B19'$ NiTi predicted with the PBE-trained model (dark blue), with PBE DFT (light blue), and with the MEAM model of Ref.\ \cite{ko2015development}.}
	\label{fig3}
\end{figure*}

\begin{figure*}
	\centering
    \includegraphics{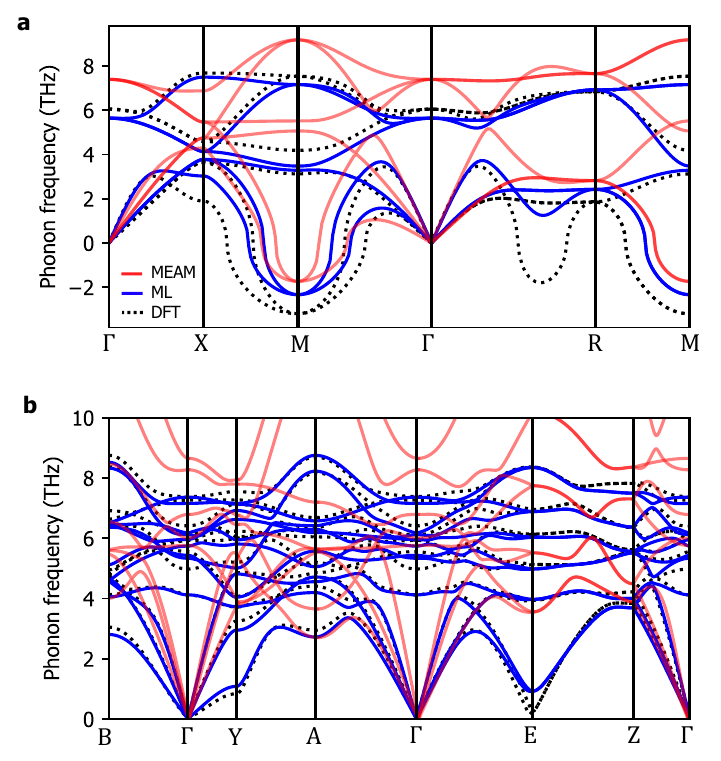}
	\caption{$B2$ and $B19'$ phonon frequencies predicted with the PBE-trained model (blue), with PBE DFT (dotted), and with the MEAM model of Ref.\ \cite{ko2015development}.}
	\label{fig4}
\end{figure*}
Predicting the thirteen independent elastic constants of the monoclinic $B19'$ phase represents an even more stringent test of the machine-learned force fields, and to our knowledge these constants have not been reported with any NiTi force field in the literature. The predictions of our PBE-trained model are plotted in Fig.\ 3 and compared against PBE DFT, displaying excellent agreement to within $\sim 10 \text{ GPa}$ for all thirteen elastic constants. Similar agreement is observed with the SCAN, LDA, and PBEsol models (Figs.\ S11-S13). Our models achieve higher accuracy on the $B19'$ elastic constants than on the $B2$ constants, likely because many low-temperature $B19'$ structures were sampled during the on-the-fly heating simulations. We also include in Fig.\ 3 predictions made with the MEAM model of Ref.\ \cite{ko2015development}, which was also trained on PBE reference data. While the MEAM model gives accurate predictions of four of the thirteen $B19'$ elastic constants ($C_{12}$, $C_{44}$, $C_{55}$, and $C_{45}$), it disagrees severely with DFT on the remaining nine, with errors of more than $100 \text{ GPa}$ for $C_{11}, C_{33}, C_{66},$ and $C_{16}$.

As an additional validation of the PBE-trained force field, we plot the vibrational phonon frequencies of the $B2$ and $B19'$ phases of NiTi in Fig.\ 4. Correct description of the $B2$ phonon frequencies is an interesting test of model transferability, as the model was trained on finite-temperature $B2$ structures above $300 \text{ K}$ while the phonon frequencies are derived from forces on structures only slightly perturbed from fully relaxed $B2$. We see good agreement between the ML-predicted phonons and PBE DFT (blue and black lines, Fig.\ 4(a)), with negative phonon frequencies near the M point in the Brillouin zone correctly reproduced. The MEAM model of Ref.\ \cite{ko2015development} predicts qualitatively accurate phonon frequencies relative to PBE DFT, but is less accurate than the PBE-trained ML model. We plot the phonon frequencies of $B19'$ martensite in Fig.\ 4b, and observe excellent agreement between the PBE-trained ML model and DFT throughout the Brillouin zone. The MEAM $B19'$ phonon spectrum is in severe disagreement with PBE DFT, with frequencies overestimated by more than $100\%$ for the highest-frequency optical branch (see Fig.\ S14).

\subsection{Large-scale MD}
Finally, we perform large-scale molecular dynamics simulations with the ML models to probe the structural phase transition in equiatomic NiTi supercells of $12,168$ atoms (see Fig.\ 5(a)). We first check the stability of $B19'$ martensite by performing heating simulations of pristine $B19'$ NiTi from $100$ to $1500 \text{ K}$ with a heating rate of $1 \text{ K/ps}$ (light colors, Figs.\ 5 and S15). All models predict that $B19'$ remains stable until transforming to $B2$, indicated by sharp drops in the atomic volume. The $B19' \rightarrow B2$ phase transition is observed at $750 \text{ K}$ for the LDA-trained model, $930 \text{ K}$ for SCAN, $780 \text{ K}$ for PBE, and $750 \text{ K}$ for PBEsol.

\begin{figure*}[h]
	\centering
    \includegraphics{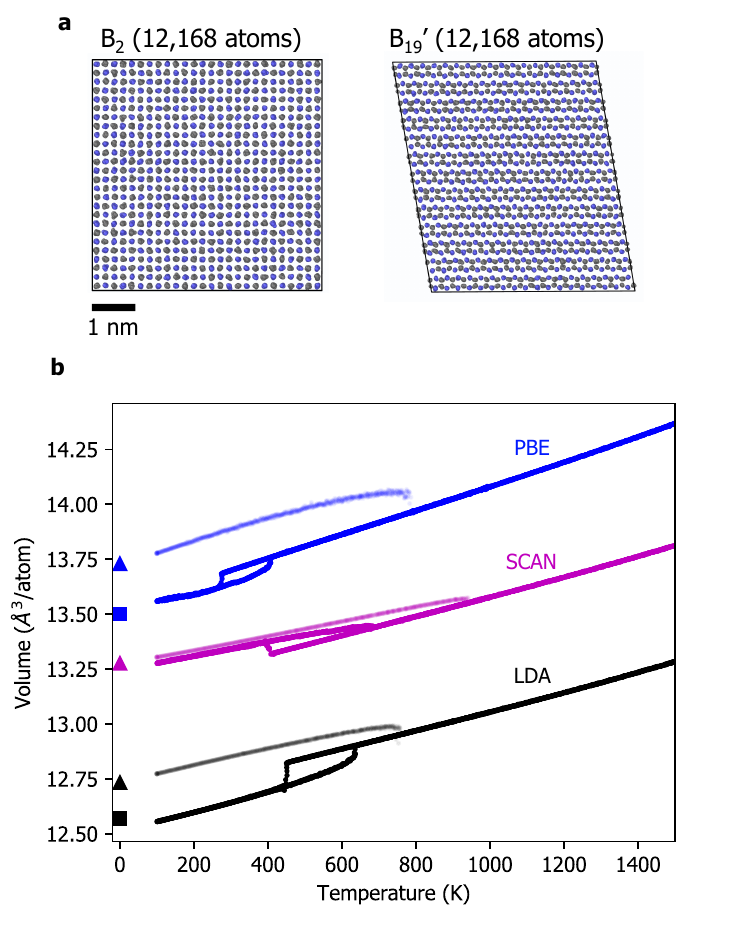}
	\caption{Large-scale MD simulations of phase transitions in equiatomic NiTi at zero pressure. (a) Snapshots of $B2$ NiTi from the SCAN cooling/heating simulation (left) and of $B19'$ NiTi from the SCAN heating simulation (right). (b) Cooling/heating simulation of $B2$ NiTi (dark) and heating simulation of $B19'$ NiTi (light colors). Zero-Kelvin volumes of $B19'$ (triangle) and $M2$ (square) are shown for reference. To avoid overlap in the plot, the PBEsol simulations are reported in Fig.\ S15.}
	\label{fig5}
\end{figure*}

Next, we perform cooling/heating simulations with each model to examine whether the $B19'\rightarrow B2$ phase transition is reversible (dark colors, Fig.\ 5(b)). For each of the four models, the simulation begins in a pristine $B2$ supercell and is equilibrated for $10 \text{ ps}$ at $1000 \text{ K}$. The system is then gradually cooled to $100 \text{ K}$ and heated to $1500 \text{ K}$ at a rate of $1 \text{ K/ps}$. All models predict reversible structural phase transitions between $\sim 200$ and $700 \text{ K}$. Consistent with the training simulations, the LDA, SCAN, and PBEsol models predict a transition to the $M2$ phase, indicated by sharp drops in the atomic volume. Only the SCAN model predicts a reversible transition to $B19'$ martensite, with the $B2 \rightarrow B19'$ transition observed at $400 \text{ K}$ and the $B19' \rightarrow B2$ transition observed at $680 \text{ K}$.

\begin{figure*}
	\centering
    \includegraphics{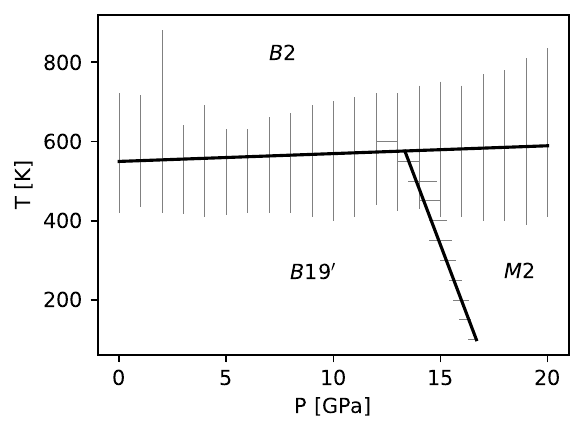}
	\caption{Phase diagram predicted by the SCAN-trained force field. Vertical gray lines indicate hysteresis windows of fixed-pressure temperature sweeps performed between 0 and 20 GPa, while horizontal lines indicate approximate phase-transition pressures derived from fixed-temperature pressure sweeps performed between 100 and 600 K (see Figs.\ S16-17 for example simulations). Estimated phase boundaries, shown in black, are derived from linear fits to the centers of the transition windows.}
	\label{fig6}
\end{figure*}

We investigate the possibility that $M2$ is stabilized at elevated pressures by performing cooling/heating and compression/decompression simulations at a range of pressures and temperatures with the SCAN-trained force field. The results of these simulations are summarized in the phase diagram shown in Fig.\ \ref{fig6}, with a clear transition to a low-volume $M2$ phase observed between 12 and 16 GPa (see Figs.\ S16 and S17 for representative plots of the system volume during these pressure and temperature sweeps). Computed XRD spectra establish that the observed low-volume phase is distinct from $B19'$ martensite (Fig.\ S18), but the spectra of this phase differ across the three force fields, indicating that the functionals agree in predicting a low volume phase but produce different atomic packing. Furthermore, symmetry analysis of frames extracted from molecular dynamics simulations of the $M2$ phase failed to reveal a space group larger than $P_1$ or $P_{-1}$, suggesting $M2$ is a low-symmetry, possibly incommensurate, phase with properties that are highly sensitive to the DFT functional.

\section{Discussion}
We have developed four force fields for equiatomic NiTi based on the LDA, PBE, PBEsol, and SCAN DFT functionals. The models were trained autonomously during molecular dynamics, without any constraints applied to the unit cell as the 144-atom model systems were heated and cooled. During these training simulations, the models are in remarkable agreement with the DFT reference to within about $1 \text{ meV/atom}$ on potential energy predictions, allowing us to generate MD trajectories at \textit{ab initio} accuracy at a fraction of the cost of \textit{ab initio} molecular dynamics. Nearly all previous computational studies of equiatomic NiTi have relied on a single DFT functional, most often PBE, and explored either the zero-Kelvin potential energy surface or the finite-temperature free energies of specific phases. Our training simulations eliminate these restrictions, allowing us to directly observe which phases form as the system is cooled or heated past the phase-transition temperature.

Surprisingly, on cooling of the $B2$ austenite phase, our LDA, PBE, and PBEsol models predict a transition to a lower-volume phase that does not match any known phases of equiatomic NiTi. We have labelled this phase $M2$. Only our SCAN model predicts the $B_{2} \rightarrow B_{19}'$ expected from experiments. Similar behavior is observed in much larger cooling and heating $12,168$-atom simulations of the phase transition, with the LDA, PBE, and PBEsol models predicting a sharp decrease in atomic volume on cooling, consistent with the zero-Kelvin volume of the $M2$ phase.

We have validated the models on a range of challenging properties, and shown excellent agreement with DFT in predictions of the elastic constants and phonon frequencies of $B2$ and $B19'$ NiTi, showing dramatic improvement over a previous MEAM force field. The accuracy of our models suggests that the departure from experiment observed at low temperature is not due to inaccuracies of the force field models in reproducing the DFT reference, but rather due to inaccuracies in DFT itself, at least with the LDA, PBE, and PBEsol functionals. This has the surprising implication that the most common DFT functionals used in the past to study NiTi, most often PBE, are in qualitative disagreement with experiment at zero pressure, predicting the more densely packed $M2$ phase rather than $B19'$ martensite. This, however, eluded detection because large-scale DFT-accurate simulations were not possible before the possibility to develop machine learning force fields.

It is possible that the $M2$ phase predicted by our LDA, PBE, and PBEsol models is stabilized at higher pressure. The phase diagram of equiatomic NiTi has not been extensively explored in experiments. A recent DFT study has proposed candidate phases of NiTi at high pressures \cite{liu2018high}, but the analysis was based on zero-temperature enthalpies and therefore excluded entropic effects. Because $M2$ is more densely packed than $B19'$, it may become favored as pressure increases. We intend to analyze both the enthalpies and finite-temperature stability of the $M2$ phase at elevated pressures, to investigate whether $M2$ may form in experiments in the high-pressure limit.

\section{Methods}

\subsection{Sparse Gaussian process models}
The models discussed in this work are sparse Gaussian processes trained with the FLARE++ code, as discussed in greater detail in Ref.\ \cite{vandermause2021active}. Here, we briefly summarize the key features of the SGP model and report our choice of hyperparameters. The SGP predicts the local energy $\varepsilon$ of a local environment $\rho_i$ centered at atom $i$ by summing over a set of representative local environments $S$,
\begin{equation}
    \varepsilon(\rho_i) = \sum_{s \in S} k(\mathbf{d}_i, \mathbf{d}_s) \alpha_s,
\end{equation}
where $k$ is a kernel function quantifying the similarity of two inputs, $\mathbf{d}_i$ and $\mathbf{d}_s$ are descriptors of the local environments $\rho_i$ and $\rho_s$, and $\boldsymbol{\alpha}$ is a vector of training coefficients. Atomic forces and virial stresses are computed from gradients of local energy predictions with respect to the atomic positions and cell vectors. In this work, we use a normalized dot product kernel with integer power $\xi = 2$,
\begin{equation}
k(\mathbf{d}_1, \mathbf{d}_2) = \sigma^2 \left( \frac{\mathbf{d}_1 \cdot \mathbf{d}_2}{d_1 d_2} \right)^\xi.
\end{equation}
The signal variance $\sigma^2$ is a continuous hyperparameter that is optimized together with the energy, force and stress noise parameters $\sigma_E, \sigma_F$ and $\sigma_S$ by maximizing the log marginal likelihood of the SGP. For the descriptor vectors $\mathbf{d}_i$, we use the $B_2$ term of the many-body Atomic Cluster Expansion \cite{drautz2019atomic} with $N_{\text{rad}} = 8$ Chebyshev radial basis functions, $\ell_{\text{max}} = 3$ in the spherical harmonics expansion, and a quadratic cutoff with radius $r_{\text{cut}} = 5 \text{ \AA}$.

Uncertainties in the local energy $\varepsilon(\rho_i)$ are computed as the square root of a normalized predictive variance $V_{\epsilon}$, defined as
\begin{equation}
V_\varepsilon = 1  - \frac{1}{\sigma^2} \mathbf{k}_{\varepsilon S} \mathbf{K}_{SS}^{-1} \mathbf{k}_{\varepsilon S}^\intercal,
\end{equation}
where $\mathbf{k}_{\varepsilon S}$ is the vector of kernels between $\rho_i$ and the environments $\rho_s$ in the sparse set, and $\mathbf{K}_{SS}$ is the matrix of kernels between pairs of sparse environments.

Final models for each DFT functional are constructed by pooling together the structures collected during the heating and cooling training simulations and training a single SGP on all structures. The optimized hyperparameters $\sigma$, $\sigma_E$, $\sigma_F$, and $\sigma_S$ from the heating simulations are used in the final trained models. The final models are trained on cohesive energies rather than potential energies by correcting raw potential energy labels with isolated-atom energies of Ni and Ti. To accelerate predictions, the final models are mapped onto equivalent parametric models that are quadratic in the descriptor $\mathbf{d}_i$ and that have a prediction cost that is independent of the size of the sparse set \cite{vandermause2021active}.

\subsection{On-the-fly training}
Training simulations are performed in the NPT ensemble using the Atomic Simulation Environment (ASE) \cite{larsen2017atomic}. We use a timestep of $5 \text{ fs}$ and characteristic thermostat and barostat timescales of $25$ and $75 \text{ fs}$, respectively. Each training simulation is initialized with a $144$-atom supercell of $B2$ or $B19'$ NiTi with the experimentally measured lattice parameters from Refs.\ \cite{goo1985b2, kudoh1985crystal}. Atomic positions of these initial structures are randomly jittered by up to $0.1 \text{ \AA}$ to give nonzero forces on the first frame. The temperature schedule for the training simulations is reported in Table I.

\begin{table}
\begin{center}
\begin{tabularx}{0.8\textwidth}{
   >{\centering}X
   >{\centering}X
   >{\centering\arraybackslash}X }
 \hline
 \hline
  Time $t$ (ps) & Cooling $T$ (K) & Heating $T$ (K) \\
 \hline
 $0$ & 1000 & 250 \\
 $5$ & 750 & 500 \\
 $10$ & 500 & 750 \\
 $15$ & 250 & 1000 \\
 $20$ & 100 & 1250 \\
 $25$ & 50 & 1500 \\
 \hline
\end{tabularx}
\label{training_stats}
\end{center}
\caption{Temperature schedule for the $50$-ps on-the-fly training simulations.}
\end{table}

During the training simulations, hyperparameter optimization of $\sigma, \sigma_E, \sigma_F,$ and $\sigma_S$ is performed with the L-BFGS-B algorithm implemented in SciPy \cite{2020SciPy-NMeth} with the energy noise $\sigma_E$ constrained to be greater than $1 \text{ meV/atom}$. Hyperparameter optimization is performed after each of the first $N_\text{hyp} = 10$ updates to the training set, after which the hyperparameters are held fixed. The sparse set of the SGP is initialized with $7$ arbitrarily chosen environments from the first frame. DFT is called whenever the predictive uncertainty on a local energy prediction exceeds $1\%$. After the DFT calculation is performed, the potential energy, atomic forces, and six independent components of the virial stress tensor are added to the training set of the SGP, and local environments with predictive uncertainty greater than $0.5\%$ are added to the sparse set. After each update to the model, a minimum of $10$ MD steps are taken with the SGP before the next DFT calculation is performed.

\subsection{DFT details}
DFT calculations are performed with plane-wave basis sets and the projector augmented-wave (PAW) method implemented in the Vienna \textit{Ab Initio} Simulation Package (VASP) \cite{Kresse93p558}. We use a plane-wave kinetic energy cutoff of $336.9 \text{ eV}$ and a k-point mesh of $\sim 4000$ k points per reciprocal atom. The Methfessel-Paxton smearing method is used with a broadening value of $0.2 \text{ eV}$. The $3\text{d}^9 4\text{s}^1$ and $3\text{d}^3 4\text{s}^1$ electrons are included in the valence for Ni and Ti, respectively. For elastic constant and phonon frequency calculations, an electronic energy convergence criteria of $10^{-7} \text{ eV}$ is used. Phonon frequencies are calculated using the frozen phonon method as implemented in Phonopy \cite{togo2015first}. For on-the-fly DFT calculations, the default convergence criteria of $10^{-4}\text{ eV}$ is used. Except for isolated-atom calculations for Ni and Ti, the calculations are not spin polarized. For the isolated-atom calculations, a vacuum size of $8 \text{ \AA}$ was used.

\subsection{MD details}
Large-scale MD simulations are performed with LAMMPS using a custom pair style available in the FLARE++ repository. Simulations are performed in the NPT ensemble at zero pressure. We use a timestep of $2 \text{ fs}$, a temperature damping parameter of $0.2 \text{ ps}$, and a stress damping parameter of $2 \text{ ps}$ in all simulations. Atomic volumes reported in Fig.\ 5 are averaged over $500$ consecutive MD steps.

\section{Acknowledgments}
This work was supported by the National Science Foundation
under Award No. DMR-1808162 and the Harvard University Materials Research Science and Engineering Center Grant No. DMR-2011754.

\clearpage
\bibliography{bib.bib}

\end{document}